\documentclass{article}

\usepackage[english]{babel}

\usepackage[letterpaper,top=2cm,bottom=2cm,left=3cm,right=3cm,marginparwidth=1.75cm]{geometry}

\usepackage{amsmath}
\usepackage{graphicx}
\usepackage[square,numbers]{natbib}

\usepackage[colorlinks=true, allcolors=blue]{hyperref}
\usepackage{lipsum}
\usepackage{url}
\usepackage{hyperref}
\usepackage{multirow}
\usepackage{caption}
\usepackage{subcaption}
\usepackage{rotating}
\usepackage{titlesec}
\usepackage{scrextend}
\usepackage{lineno}
\titleformat{\paragraph}
{\normalfont\normalsize\bfseries}{\theparagraph}{1em}{}
\titlespacing*{\paragraph}
{0pt}{3.25ex plus 1ex minus .2ex}{1.5ex plus .2ex}
\title{Resistive Plate Chambers and Gaseous Detector Activities in the Brazilian High Energy Physics Community}

\author{Organizers: Sandro Fonseca de Souza\footnote{Rio de Janeiro State University - UERJ - \texttt{sfonseca@uerj.br}} , Gilvan Augusto Alves\footnote{Brazilian Center for Physics Research - CBPF - \texttt{gilvan@cbpf.br}} \\
\\ Authors: Mapse Barroso (UERJ), Helio Nogima (UERJ), Felipe Silva (UERJ and UEA),\\João Pedro Gomes Pinheiro (UERJ), Katherine Maslova Defante (UERJ and CERN),\\ Dalmo Dalto (UERJ), Mauricio Thiel (UERJ), Luís Mendes (CBPF/LIP)}

\begin{document}
\maketitle

\begin{abstract}
This document presents an overview of the activities of Brazilian groups on gaseous detectors, with emphasis on resistive plate chambers (RPCs) for the CMS experiment at the CERN LHC and related applications. It summarises the design, performance, maintenance and upgrade of the CMS RPC system, including the installation of improved RPCs for the HL-LHC era and the development of alternative, lower-GWP gas mixtures. It also describes local laboratory infrastructures in Brazil for RPC assembly, testing and gas studies, as well as spin-off projects such as the MARTA cosmic-ray detector and applications in medical imaging and muography. It is intended as a living reference to be periodically updated, documenting the contributions of the Brazilian community to RPC technology and its applications.
\end{abstract}
\tableofcontents
\newpage
\section{Gaseous Detectors: Introduction}
\label{section_gas}

Gas detectors operate through the ionization of a gas or a mixture of gases. A charged particle or a quantum of radiation crosses the gas and creates electron ion pairs. In the presence of a suitable electric field, the electrons drift toward the anode and can produce secondary ionization. The multiplication of electrons leads to a controlled electrical discharge that can be read out as an electrical signal. The early work of Ernest Rutherford\footnote{Chemistry Nobel Prize in 1908} on the detection of ionization in gases laid the foundations of modern gaseous detectors. During his studies of the atomic structure, he developed instrumentation able to detect individual ionization tracks left in a gas by natural radioactivity. John Sealy Townsend then investigated in detail the multiplication of charge in gases in the presence of high electric fields and described what is now called the Townsend avalanche. Hans Geiger contributed to the development of counters capable of amplifying a weak primary ionization signal, which evolved into proportional counters able to detect very small amounts of primary charge with the electronics available at the time. Geiger and his PhD student Walther M\"uller later improved these devices and made the detection of single electrons feasible.

Today gaseous detectors are still a central technology in high energy and nuclear physics experiments. They are used for precise tracking of charged particles, for timing measurements, and for triggering in many collider and fixed target experiments. Different detector families have been developed over the years, including multi wire proportional chambers, drift tubes, resistive plate chambers, and micro pattern gaseous detectors such as gas electron multipliers and Micromegas. These devices offer large sensitive areas, flexible geometries, and moderate cost, which makes them attractive for large experiments at the CERN Large Hadron Collider and at other facilities around the world \cite{Sauli_gaseous_detectors,Titov_gaseous_review}. Beyond fundamental physics, gaseous detectors are also used in medical and applied imaging. Examples include time of flight positron emission tomography scanners based on resistive plate chambers, muon radiography of volcanoes and civil engineering structures, and detector systems developed for security and industrial inspection \cite{Belli_rpc_pet,Martins_rpc_pet,Varga_muography,Muography_applications}. In this context, the goal of this document is to present an updated overview of the activities of the Brazilian community on gaseous detectors, in particular on resistive plate chambers and related technologies for the CMS experiment at the CERN LHC and for associated laboratory and spin-off applications.

\newpage



\section{CMS Experiment: Resistive Plate Chambers}
\label{section_rpc}

\subsection{Introduction}
The seminal paper on the Resistive Plate Chamber (RPC) technology was presented by R.~Santonico and R.~Cardarelli, in which they described a direct current operated particle detector whose active volume is a gas gap between two parallel bakelite electrodes~\cite{rpc_seminal}. The key idea behind the RPC, in comparison with other gaseous detectors, is the use of resistive plates as anode and cathode. The high bulk resistivity of these plates limits the size of the region where the electric field is disturbed after a discharge. As a result the detector has only a small localized dead area and can reach very good time resolution while remaining mechanically simple and relatively low cost. These features made RPCs natural candidates to cover large surfaces in many experiments.

The working principle of an RPC is based on the interaction of an ionizing particle with the gas molecules in the gap between the two plates. When a charged particle crosses the detector it creates primary electron ion pairs in the gas. Under the action of a strong and uniform electric field the electrons drift and can produce secondary ionization, giving rise to an avalanche of electrons. The gas mixture plays a central role in this process. Early RPCs used a simple argon and butane mixture~\cite{rpc_seminal}. Modern systems make use of more complex mixtures that enhance the primary ionization, control the avalanche growth, and quench secondary effects such as photon feedback and secondary discharges. An extensive review of RPC designs, performance, and applications is given in Ref.~\cite{livro_rpc}.

RPCs can operate in two main regimes, usually called streamer mode and avalanche mode. In streamer mode most of the charge multiplication occurs in the form of a streamer discharge. The induced charge is large, which allows the use of relatively simple readout electronics. On the other hand the large charge per event increases the local dead time and accelerates the aging of the detector, which limits the rate capability. For this reason streamer mode is mainly used in applications where the particle flux is low, such as cosmic ray experiments. In high rate environments like the Large Hadron Collider the RPCs are operated in avalanche mode. In this regime the discharge is strongly quenched and remains very localized. The total charge per event is smaller, which improves rate capability and aging, but requires more sensitive readout electronics to detect the signal. A broader discussion of electrical discharges in gases and of the transition between avalanche and streamer regimes can be found in Ref.~\cite{livro_descarga}.

\subsection{Description and operation of the RPCs installed in the CMS Experiment}

\begin{figure}[ht]
    \begin{center}
    \includegraphics[width=0.64\textwidth,keepaspectratio]{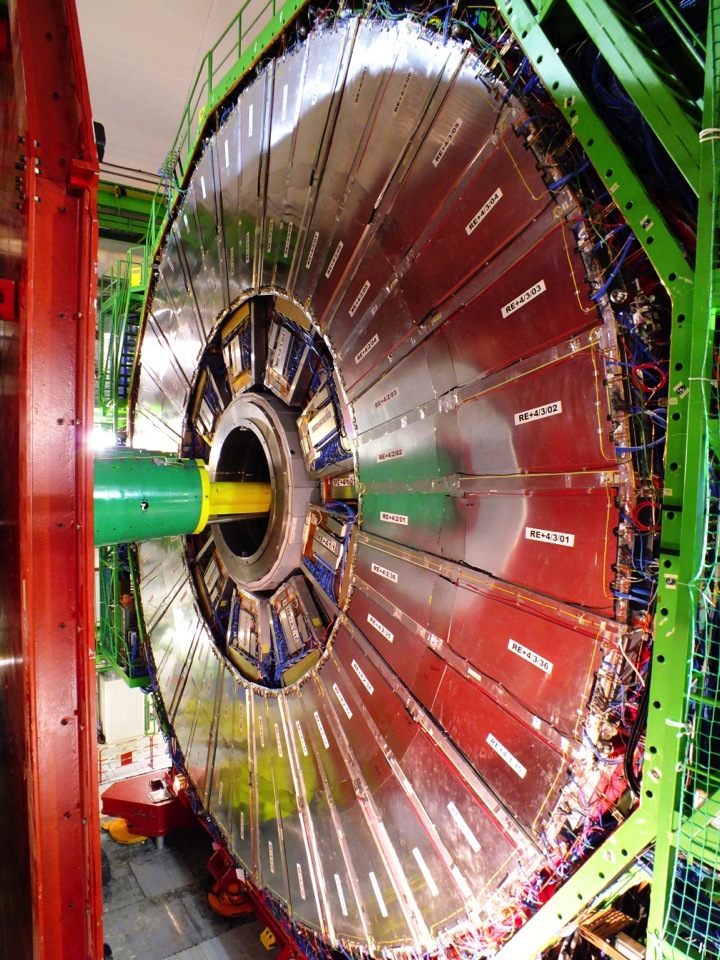}
    \end{center}
    \caption{RPC chamber installed on station RE+4 of the CMS endcap. Source:~\cite{rpc_picture}.}
    \label{picture_rpc}
\end{figure}

At CMS, the RPCs are installed in both the barrel and endcap regions and form a high performance system together with the Drift Tubes (DT) in the barrel and the Cathode Strip Chambers (CSC) in the endcap. As described in the CMS Muon Technical Design Report (Muon TDR)~\cite{muon_tdr}, the system is composed of 573 endcap chambers and 480 barrel chambers. Figure~\ref{picture_rpc} shows a picture of an RPC chamber installed on station RE+4 of the endcap.

Each chamber consists of two gas gaps (a double gap configuration). Each gap is 2~mm thick and is made of bakelite (phenolic resin) with a bulk resistivity in the range $10^{10}$ to $10^{11}\,\Omega\cdot\mathrm{m}$. The choice of electrode resistivity has a direct impact on the rate capability of the detector. The external surface of each gap is coated with a thin layer of graphite paint, which acts as a resistive electrode and distributes the applied high voltage (HV). A PET film is placed on top of the graphite for electrical insulation. A sheet carrying copper readout strips is positioned between the two gaps, and the whole structure is enclosed in an aluminium casing that provides mechanical protection and shielding.

The double gap configuration increases the detection efficiency because the signal seen by the readout strips is the logical OR of the signals from the two gaps. If only one gap is fully operational, the chamber loses about 15\% of efficiency. In this situation most of the efficiency can be recovered by increasing the applied high voltage and shifting the operating point of the chamber. The operation point is chosen in such a way that the chamber reaches the required efficiency while keeping the cluster size and the noise rate under control.

A key feature that differentiates the CMS RPCs from many previous applications of RPCs in high energy physics is the operation mode. As discussed in the previous subsection, RPCs can operate in one of two modes: streamer or avalanche. In the CMS experiment, they are run in avalanche mode. This is achieved by applying a specific high voltage and using a gas mixture that highly quenches discharges, thereby localizing the charge multiplication to a small, contained avalanche. This choice provides a rate capability of about $1~\mathrm{kHz}/\mathrm{cm}^{2}$, while streamer operation is typically limited to about $100~\mathrm{Hz}/\mathrm{cm}^{2}$. The high rate capability is essential to cope with the LHC luminosity, in particular in the regions where the background is high. In addition to rate capability, the design goals of the CMS RPC system include high efficiency (above 95\%), low cluster size (below 3 strips on average) for good spatial resolution, and good timing performance so that the signal is read out within the 25~ns of a bunch crossing and can be used by the CMS trigger system. These requirements drive the choice of materials, geometry, front end electronics and gas mixture.

\begin{figure}[ht]
\begin{center}
\includegraphics[width=1.\textwidth,keepaspectratio]{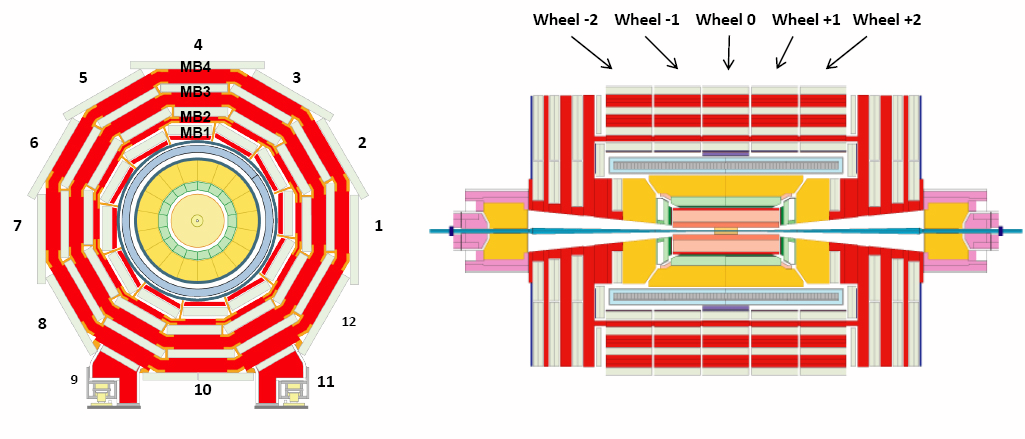}
\end{center}
\caption{$R\text{--}\phi$ (left) and $R\text{--}Z$ (right) projections of the barrel muon system.}\label{fig:barrel_rphi_rz}
\end{figure}

In the barrel region, along the radial direction, there are four muon layers called stations MB1 to MB4. In stations MB1 and MB2 a DT chamber is sandwiched between two rectangular RPC chambers, RB1 and RB2. Stations MB3 and MB4 contain a single RPC layer, RB3 and RB4. In most sectors these stations are composed of two RPC chambers, conventionally named minus and plus with increasing azimuthal angle $\phi$, which are attached to one DT chamber. In sectors 9 and 11 there is only one RPC chamber per station. One sector is a special case and hosts four chambers, labelled - -, -, + and + +. The four barrel stations are replicated along the $z$ direction in five wheels of the CMS detector (W$-2$, W$-1$, W0, W$+1$ and W$+2$) and in twelve azimuthal sectors (S1 to S12). Figure~\ref{fig:barrel_rphi_rz} illustrates the layout of the barrel muon system in the $R\text{--}\phi$ and $R\text{--}Z$ projections.

\begin{figure}[ht]
\begin{center}
\includegraphics[width=0.8\textwidth,keepaspectratio]{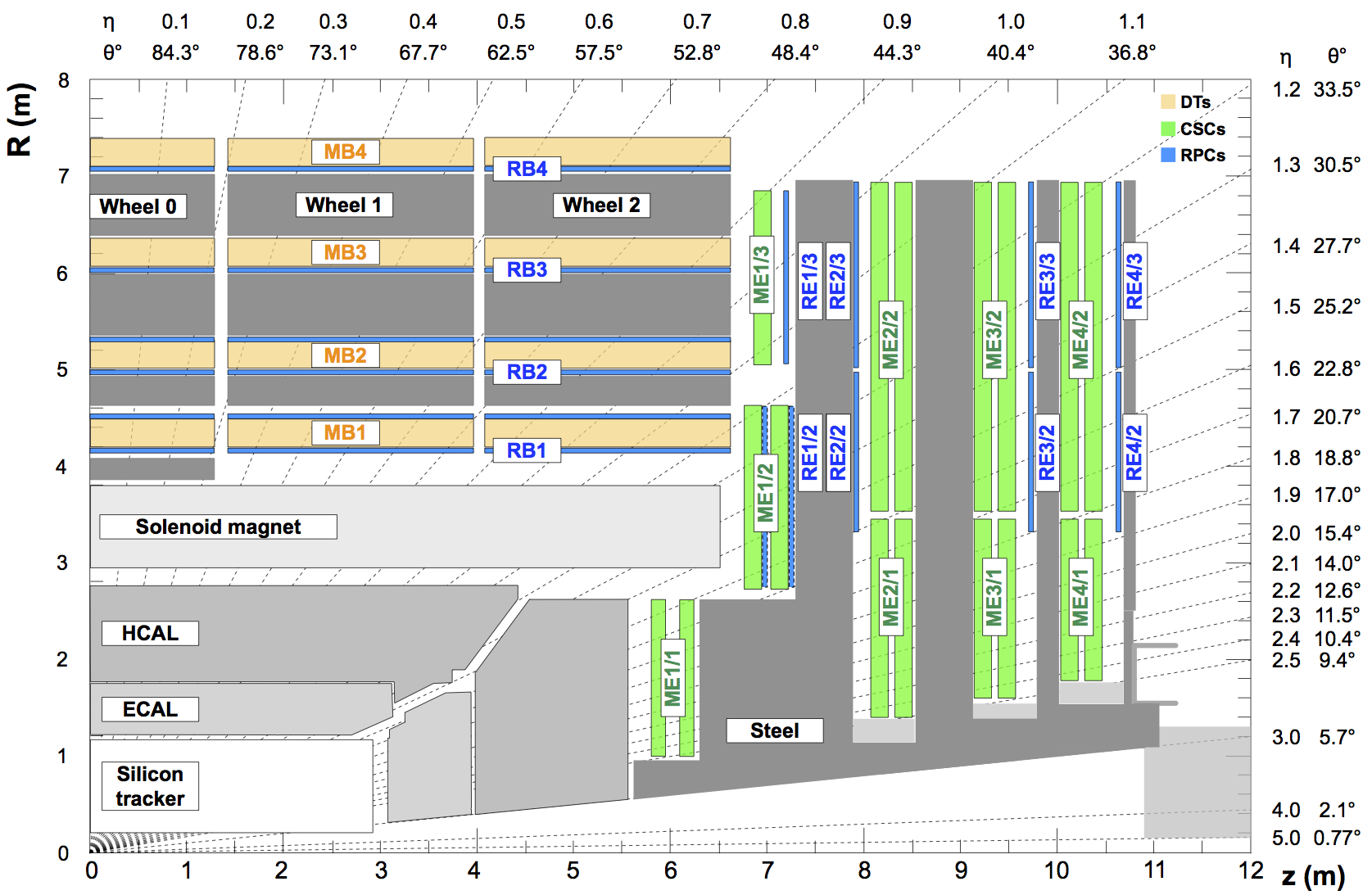}
\end{center}
\caption{$R\text{--}Z$ projection of the endcap muon system on the positive $z$ side. The same configuration is repeated for the 36 $\phi$ sectors.}\label{fig:endcap_rz}
\end{figure}

In the endcap region, the RPC chambers have a trapezoidal shape and are distributed in four disks, or stations, on each side of the detector, named RE$\pm1$, RE$\pm2$, RE$\pm3$ and RE$\pm4$. Each disk is divided into 72 chambers. The disks are segmented into three concentric rings along the radial direction and 36 sectors in azimuthal angle. RPCs are installed in the two outer rings, R2 and R3, in all 36 sectors. The RE$\pm4$ disks are a special case because their chambers were installed only in 2014. For these disks a design choice was made to mechanically attach the R2 and R3 chambers of each sector into a single structure, called a super module. Figure~\ref{fig:endcap_rz} shows the $R\text{--}Z$ projection of the endcap muon system for the positive $z$ side. The same configuration is repeated in all 36 $\phi$ sectors.

The strip length is chosen, in both barrel and endcap, to control the area of each strip and to reduce fake muons due to random coincidences. The design takes into account the time of flight of muons and the signal propagation along the strip. In the barrel each chamber is segmented longitudinally into two readout regions, called rolls, named forward and backward along increasing $|\eta|$.\footnote{Some chambers are divided into three rolls, forward, middle and backward, for trigger purposes.} In the endcap the strips are segmented into three radial regions, called partitions A, B and C, from the inner to the outer radius of the disk.

The gas mixture used in the CMS RPCs is composed of C$_2$H$_2$F$_4$ (Freon R134a, tetrafluoroethane), C$_4$H$_{10}$ (isobutane) and SF$_6$ (sulphur hexafluoride) in a 95.2:4.5:0.3 ratio, with a controlled relative humidity of about 40\% at 20--22~$^{\circ}$C. The Freon component provides the main ionization and charge multiplication that characterise the avalanche. The isobutane acts as a quencher, reducing secondary ionizations and suppressing the formation of streamers. The SF$_6$ is added in small concentration to further suppress discharges and reduce spurious background signals. The choice of a Freon based mixture over argon or helium based alternatives was guided by dedicated studies performed during the R\&D phase~\cite{gas_mixture_BERNARDINI1995428,gas_mixture_Gorini}.

Since the early R\&D phase, RPCs have shown good performance and stability under irradiation and over long operating periods. This behaviour was first observed in previous RPC based experiments~\cite{Bressi:1987kg,Ge:2014tea,Abbrescia:1993xy,Antoniazzi:1992dg,DiCiaccio:1992di,deAsmundis:1995dq,Boutigny:1995ib}. More recent aging studies, performed under conditions that reproduce and even exceed those expected at the High Luminosity LHC, with a safety factor of about three on the background rate (up to 600~Hz/cm$^{2}$), confirm the robustness of the CMS RPC design and its suitability for long term operation~\cite{andrea_rpc_2018}.

\newpage

\subsection{Performance}

For the CMS experiment, the present RPC system contributes mainly to the muon trigger, thanks to its very good time resolution. The two main quantities used to monitor its performance are the efficiency and the cluster size. The efficiency is defined as the ratio between the number of registered RPC hits and the number of muons that cross a given detector unit. The cluster size is defined as the number of adjacent readout strips fired by a single muon hit. Both quantities are measured regularly during data taking and in dedicated calibration runs.

In normal LHC running conditions the CMS RPC system operates with an average efficiency above 95\% and an average cluster size smaller than 3, which were the target values defined during the design phase. The efficiency is the most intuitive parameter, since it directly tells how often a muon produces a visible signal in the detector. The cluster size is less obvious but equally important. As discussed before, the RPCs in CMS are operated in avalanche mode, where the electrical discharge remains confined in a small region of the gas gap. This limits the induced signal to a small group of strips. If the detector were to work in streamer mode, secondary discharges and photon feedback would fire more neighbouring strips and the cluster size would grow.

Keeping the mean cluster size under control, below about three strips, has two direct benefits. First, it improves the spatial resolution, because the hit position used in the reconstruction is taken as the centre of the fired cluster. Second, it protects the system from saturation effects. A very sensitive front end electronics is required to detect the small signals produced in avalanche mode. If many strips fire at the same time, or if noise and dark counts are high, the electronics can suffer from dead time and pile up, reducing the useful rate capability of the detector. A detailed review of the CMS RPC performance during Run 2 is given in Ref.~\cite{rpc_run2_performance}.

The working point calibration is the procedure used to choose the high voltage (HV) at which each RPC chamber is operated. For a given gas mixture and set of environmental conditions, the efficiency and the cluster size of a chamber depend on the effective HV. During a calibration scan the applied voltage is varied in steps and, at each point, the efficiency and cluster size are measured. These curves are then used to select a working point in the plateau region, where the efficiency is high and stable and the cluster size remains below the target value. The procedure also takes into account changes in temperature and pressure, which modify the effective HV, and possible aging effects in the detector. Regular working point calibrations are therefore essential to keep the performance of all chambers uniform in time and to ensure that the RPC system continues to operate safely and efficiently as the LHC conditions evolve. Recently, a machine-learning-based analysis of the HV scan data has been introduced to extract the working points in an automated and more robust way~\cite{rpc_hvscan_ml}.

\begin{figure}
    \centering
    \includegraphics[width=1.\linewidth]{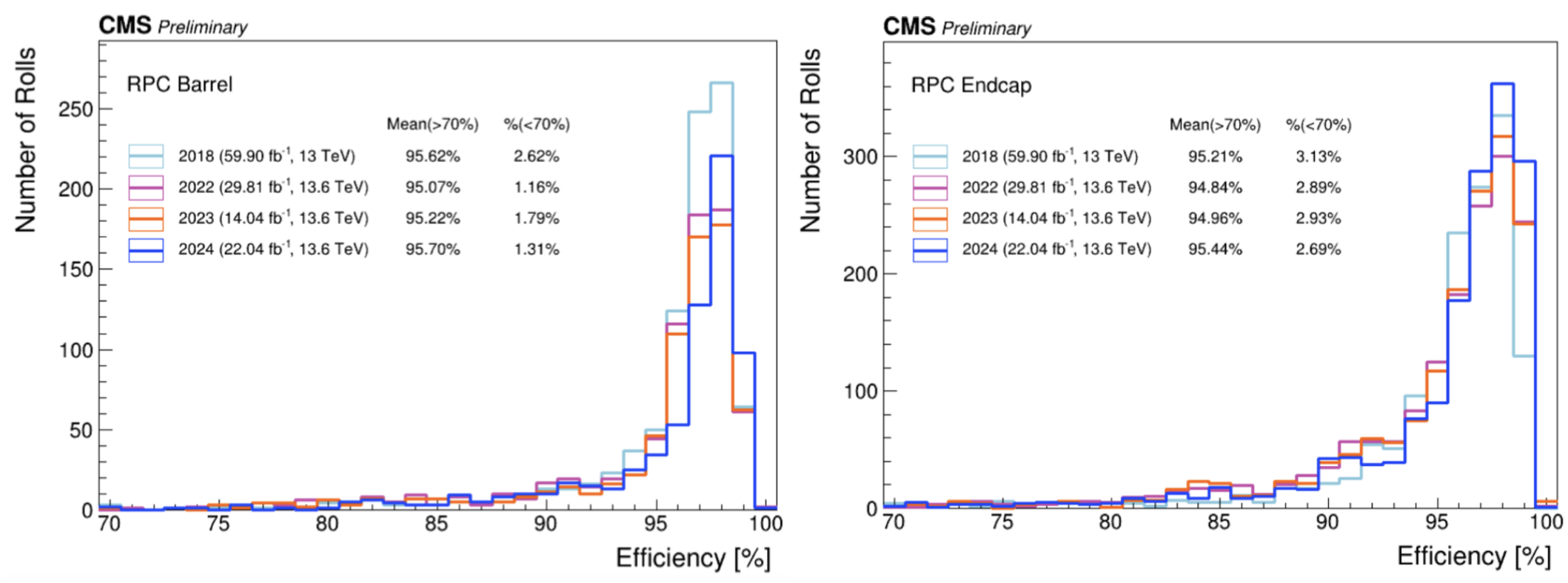}
    \caption{RPC overall efficiency distribution comparison for the barrel (left) and endcap (right) regions, obtained using 2018, 2022, 2023 and 2024 proton-proton collisions data.}
    \label{fig:RPCEffRun3}
\end{figure}

The CMS experiment has already collected more than 180~fb$^{-1}$ of integrated luminosity in Run~3, at a centre-of-mass energy of 13.6~TeV. At the beginning of each data-taking year a dedicated HV scan is performed at an instantaneous luminosity of about $2\times 10^{34}\,\text{cm}^{-2}\text{s}^{-1}$, with data taken at several HV values in the range 8.6--9.8~kV, in order to determine the working points for all RPC chambers~\cite{rpc_hvscan_ml}. Under these conditions, and during regular physics runs at similar or higher luminosities, the RPC system continues to show an average efficiency of about 95\% in both barrel and endcap regions, as shown in Figure~\ref{fig:RPCEffRun3}, with an average cluster size below 2, as already observed in Run~2. This level of performance is the result of the annual HV scans, continuous monitoring of the system, and dedicated maintenance during long shutdowns and year-end technical stops. The calibration and operation activities are coordinated by the RPC operation team and the RPC Technical Coordination. These groups follow the online monitoring in real time and can intervene quickly whenever a problem is identified, for example by recovering chambers in single-gap operation, repairing gas leaks when access is possible, and updating the working points derived from each yearly scan. A recent summary of the Run~3 operation and performance, including the impact of the new freon recuperation system and of the large-scale gas leak repair campaign, can be found in Ref.~\cite{rpc_run3_performance}.





\subsection{RPC maintenance}
\label{RPCMaintenance}
During LHC running periods the RPC system is continuously monitored and maintenance actions are implemented whenever access allows, while more extensive interventions are carried out during access periods in Year End Technical Stops (YETS) and long shutdowns (LS). The goal is to keep all chambers safe and efficient and to prepare the detector for future running conditions. Typical interventions address problems in the high voltage (HV) and low voltage (LV) distribution, in the front end electronics and connectors, in the gas system and in the slow control. The work that can be done depends on the level of access to the detector. Components installed in the service cavern can be reached at any time. Replacement of LV boards or Link boards can be done during short beam-dump interventions, and interventions on the chambers themselves require CMS to be opened in a specific configuration during a YETS or an LS.

The HV power supplies are installed in the service cavern. Faulty HV boards and their crate side connectors can therefore be replaced whenever needed. The LV boards for the RPC front end electronics are mounted on the towers in the experimental cavern and can only be accessed when the cavern is open to personnel (no beam). The same distinction holds for connectors. If a problem is located at the board side it can be solved together with the board replacement. If the damaged connector is on the detector side, close to the chamber, access is only possible when CMS is opened in specific configuration during a YETS or an LS. Replacing a LV or signal connector at the chamber usually takes a few hours, while repairing or replacing a HV connector is more delicate and typically requires two or three days, including checks at low and nominal voltage. During LS2, about sixty HV channels and a dozen LV channels were recovered in this way, leading to a clear increase in the average efficiency of the system~\cite{rpc_ls2_jinst,rpc_ls2_lhcp}.

Problems in the front end boards (FEBs) also require direct access to the chambers. In the barrel the chamber must be partially or totally extracted from the yoke to reach the FEBs, which usually takes one or two days when preparation, safety checks and recommissioning are included. In the endcap the FEBs are behind removable covers and can be replaced in a few hours once access is granted, but CMS should be open in a specific configuration. These interventions are often combined with updates of the LV distribution, threshold tuning and noise scans, so that the recovered channels can be brought back to stable operation before the detector is closed again~\cite{rpc_ls2_jinst}.

\begin{figure}
    \centering
    \includegraphics[width=0.8\linewidth]{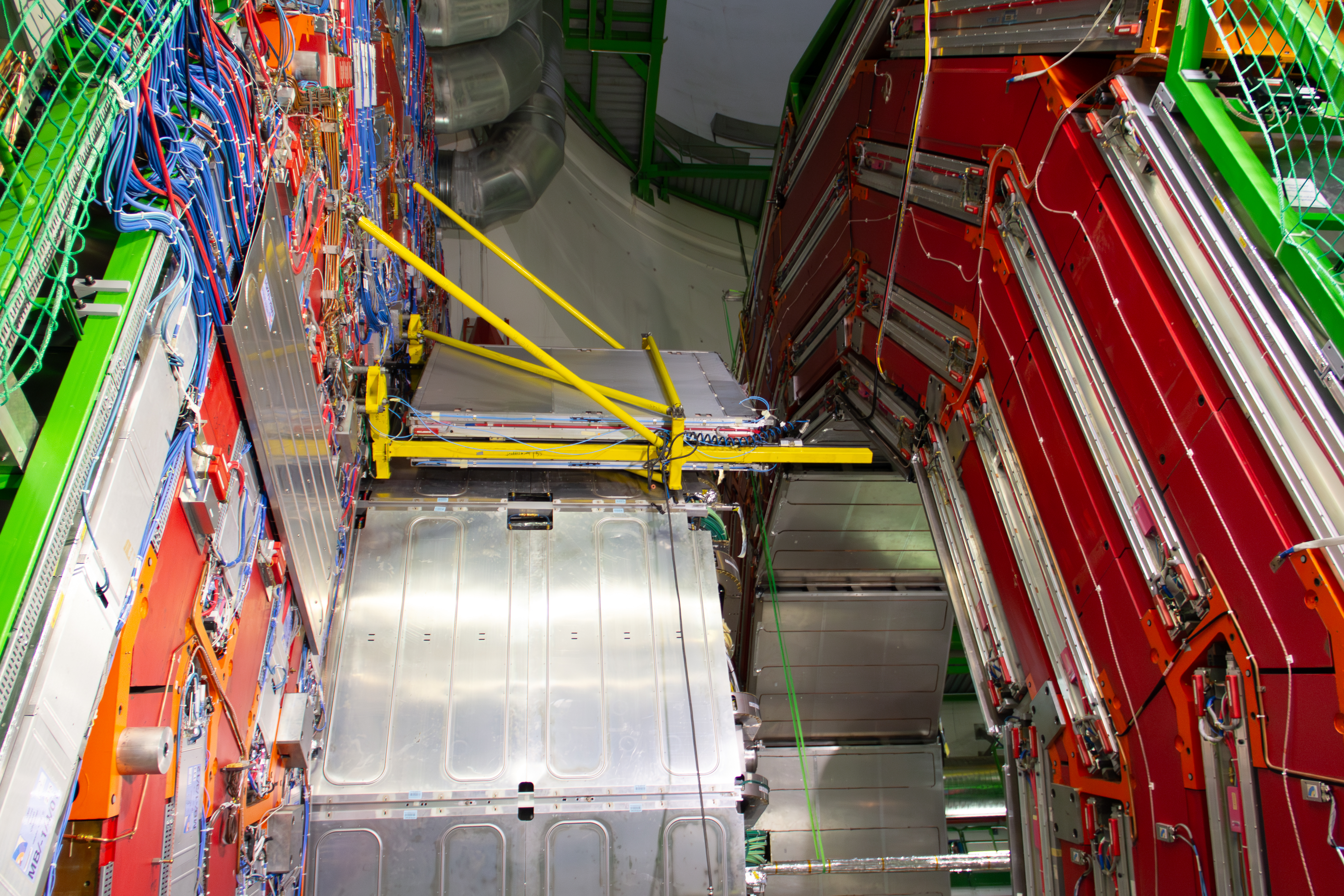}
    \caption{CMS RPC barrel chamber fully extracted from yoke during the new leak repair campaign in YETS\,24/25.}
    \label{fig:photoleakrepair}
\end{figure}

Gas leaks are another common issue, particularly in barrel chambers. They lead to an increased emission of the gas mixture and therefore to higher operational costs and larger emissions of greenhouse gases. During LS2 a dedicated leak repair campaign was carried out. Chambers with significant leaks were partially extracted and the leaking sections of the inlet or outlet pipes were repaired locally. About half of the leaky barrel chambers could be recovered with this method, resulting in a measurable gain in average efficiency and in a reduction of the total leak rate~\cite{rpc_ls2_jinst,rpc_ls2_lhcp}. Based on the experience from LS2, a new and more robust strategy was adopted starting from the YETS 2023/2024. In this procedure the affected chamber is fully extracted, as shown in Figure\ref{fig:photoleakrepair}, and all gas pipes, from the inlet to the patch panel, are replaced. Although this approach is more time-consuming, since it requires full chamber extraction and pipe replacement, it provides a definitive repair. Chambers treated with the full extraction procedure have not developed new leaks and the overall leak rate of the system has been significantly reduced~\cite{rpc_run3_performance}.

In parallel to local repairs, the LS2 program included a consolidation of the central gas system. Automatic pressure regulation was introduced to reduce mechanical stress on the gaps and to stabilise the operating conditions, and the first C$_2$H$_2$F$_4$ recuperation system with an efficiency of about 80\% was installed. These measures reduce the environmental impact of the detector and help protect the chambers during pressure excursions~\cite{rpc_ls2_jinst,rpc_ls2_lhcp}.

The intervention on the RE4 stations is an example of large scale maintenance. All 72 supermodules of the fourth endcap station were extracted to the surface to allow access to the CSC ME4/1 chambers. A dedicated laboratory with controlled temperature and humidity was set up at CMS Point~5 to host the supermodules. There the RPC chambers were powered and commissioned with cosmic rays. The currents measured at the beginning of the conditioning were higher than at the end of Run~2, but they decreased with time at high voltage until they reached values compatible with their previous history. Only one chamber had to be replaced by a spare unit, while all others were certified for reinstallation in CMS after detailed checks of noise, front end electronics and gas tightness~\cite{rpc_ls2_jinst,rpc_ls2_lhcp}. Once the hardware work is completed, the recovered chambers are reintegrated into standard operation with updated working points, online monitoring and the first collision runs. In this way, the maintenance program carried out during LS2, the following YETS and later shutdowns has been essential to keep the RPC system efficient and reliable for Run~3 and for the preparation of the Phase~2 upgrade~\cite{rpc_run3_performance}.

\subsection{Improved RPC chambers for HL-LHC}
\label{section:iRPC}

\begin{figure}[ht] \begin{center} \includegraphics[width=.9\textwidth,keepaspectratio]{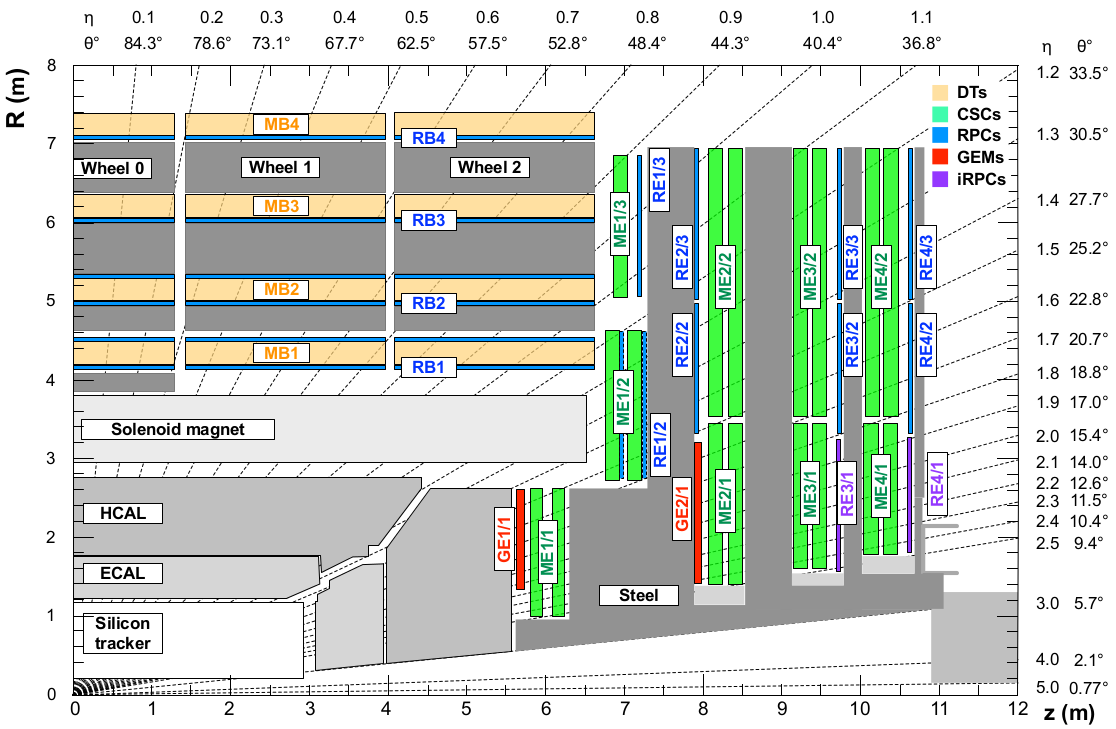} \end{center} \caption{$\eta$ projection of the Muon System subdetectors. In purple, is labeled the iRPCS to be installed during the CMS upgrade.}\label{muons_eta} \end{figure}

During the Phase--2 of the LHC, known as the High Luminosity LHC (HL-LHC), the instantaneous luminosity is expected to reach about $5\times 10^{34}\,\text{cm}^{-2}\text{s}^{-1}$ and CMS is projected to collect an integrated luminosity of order 3000~fb$^{-1}$. In these conditions the muon system must preserve good trigger performance and efficient reconstruction, in particular in the forward region. The present RPC system covers the barrel and the outer part of the endcaps up to $|\eta|\approx 1.9$, while the very forward region $1.9<|\eta|<2.4$ is instrumented only with CSCs. To provide additional readout and improve the trigger and particle reconstruction in this region, the CMS Phase--2 Muon Upgrade foresees the installation of new improved Resistive Plate Chambers (iRPCs) on the inner rings of the third and fourth endcap stations, RE3/1 and RE4/1~\cite{muon_tdr,improved_rpc_hllhc}, as shown in Figure~\ref{muons_eta}.

In the HL-LHC scenario the iRPCs in RE3/1 and RE4/1 are expected to operate at average background rates of a few hundred Hz/cm$^{2}$, with local maxima close to 700~Hz/cm$^{2}$ and design values, including a safety factor of three, reaching about 2~kHz/cm$^{2}$~\cite{rpc_background_hllhc}. To cope with these conditions the iRPCs adopt a thinner gas gap of 1.4~mm and thinner High Pressure Laminate (HPL) electrodes compared to the present RPCs. The reduced gap lowers the charge per avalanche and improves the rate capability, while the choice of electrode material preserves a good timing performance. A new front end electronics with a threshold of about 30~fC and time resolution better than 150~ps reads out strips from both ends, which allows a precise measurement of the hit position along the strip and a time resolution at the chamber level of order 1.6~ns~\cite{improved_rpc_hllhc}. Beam and irradiation tests at GIF++ have shown plateau efficiencies above 95\,\% with an average cluster size around two strips in the expected rate range, confirming that the iRPC design satisfies the HL-LHC requirements~\cite{improved_rpc_hllhc}.

Mass production of iRPC chambers is carried out at CERN and at an external assembly site in Ghent, following a common assembly and quality control procedure. Each chamber is built from two single gaps with HPL electrodes, graphite coating and honeycomb mechanical support, and is equipped with readout strips and new generation front end boards. The quality control chain includes gas tightness and high voltage conditioning tests, efficiency and cluster size measurements with cosmic rays at several high voltage values and visual inspection of the mechanical and electrical components. Only chambers that pass all these steps are accepted for installation in CMS. The Brazilian group has played a central role in the definition and execution of these procedures, both in the surface laboratories and during CMS installation and commissioning.

\begin{figure}
    \centering
    \includegraphics[width=0.9\linewidth]{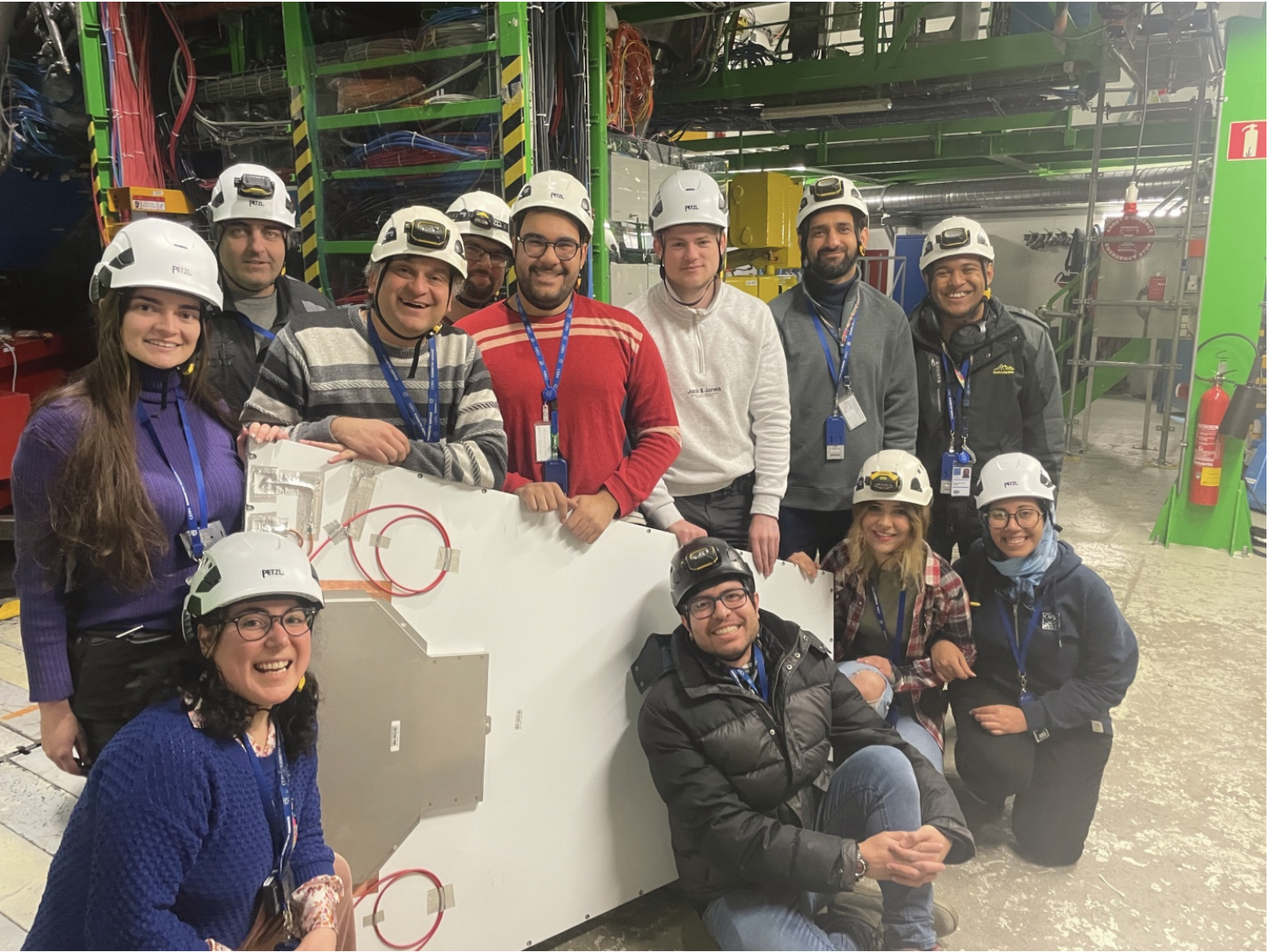}
    \caption{iRPC installation team with the first iRPC mass production chamber installed in January 2025 at CMS experimental cavern.}
    \label{fig:RPCinst}
\end{figure}

The installation of the iRPCs on the detector started with a demonstrator phase, in which a small number of final design chambers were installed and operated in CMS to validate the mechanics, services and electronics under real conditions~\cite{cms_muon_upgrade_cr2024}. After the successful completion of this phase, the first large installation campaign took place during the LHC YETS2024/2025. In this period 36 sectors of iRPC chambers were installed on the negative endcap, covering the inner rings of the RE$-3/1$ and RE$-4/1$ stations. Each of these rings hosts 36 trapezoidal chambers, one per azimuthal sector, so this campaign delivered the first half of the final iRPC system~\cite{improved_rpc_hllhc,irpc_news}. The work required opening the space between the endcap disks, draining and reconnecting the cooling circuits, installing new gas and power services, mounting the chambers on their supports and performing leak checks, high voltage tests and data taking runs in the underground cavern~\cite{irpc_news}. The installation was completed in record time thanks to the detailed preparation and to the strong coordination between the RPC Technical Coordination, the Run coordination and the installation crew, which included several members of the Brazilian group, as shown in Figure~\ref{fig:RPCinst}. The corresponding iRPC chambers for the positive endcap are planned to be installed at the beginning of LS3, completing the Phase--2 upgrade of the RPC system.

\subsection{Alternative gas mixtures for Resistive Plate Chambers -- RPC}
\label{subsec:alt_gas_mixtures}

One important constraint for the operation of large RPC systems at the HL-LHC is the environmental impact of the working gas. The standard CMS-RPC mixture is dominated by tetrafluoroethane (C$_2$H$_2$F$_4$), a fluorinated gas with a high global-warming potential (GWP). When the contributions of all components are combined, the effective GWP of the mixture is about 1500, so that releasing 1~kg of gas has the same greenhouse effect as about 1.5~t of CO$_2$~\cite{bib:Pinheiro2024_CO2Mixture_NIMA}. At the same time, the European Green Deal sets the goal of reducing net greenhouse-gas emissions by at least 55\% by 2030 and of reaching climate neutrality by 2050~\cite{bib:EU_Green_Deal}. At CERN, about 90\% of direct (Scope~1) emissions come from experimental activities, and more than three quarters of these are due to fluorinated gases used in refrigeration systems and detector mixtures~\cite{bib:CERN_EnvReport_2021_2022_Emissions}. The CERN F-gas Policy adopted in 2024 therefore calls for a progressive reduction of fluorinated-gas consumption and emissions, including a decrease of direct emissions by about 28\% by the end of Run~3 with respect to 2018~\cite{bib:CERN_FGAS_Policy_2024}.

In the CMS-RPC system several measures are already in place to limit gas consumption and accidental releases. The detector operates in closed-loop recirculation, a C$_2$H$_2$F$_4$ recuperation plant with an efficiency of about 80\% has been installed and commissioned~\cite{bib:Pinheiro2024_CO2Mixture_NIMA}, and large repair campaigns are ongoing to fix leaky gas lines and connectors, as shown in Section~\ref{RPCMaintenance}. These actions substantially reduce the demand for fresh C$_2$H$_2$F$_4$ and avoid part of the potential emissions to the atmosphere. They do not, however, change the fact that the working mixture itself still has a high GWP. To reach the long-term climate targets, it is therefore also necessary to act on the gas composition.

Within this context the RPC community, both in CMS and in other experiments, is developing alternative gas mixtures that reduce the GWP at the source while preserving the detector performance required for LHC operation. Two complementary strategies are being followed. A long-term approach aims at a full replacement of C$_2$H$_2$F$_4$ by hydrofluoroolefin (HFO) gases, in particular C$_3$H$_2$F$_4$ (HFO-1234ze), often combined with CO$_2$, so as to reach mixture GWPs of a few hundred~\cite{bib:CMS_HFOmixtures_NIMA2026}. In parallel, a short- to mid-term approach keeps C$_2$H$_2$F$_4$ as the main component but replaces part of it with CO$_2$, which directly reduces the amount of high-GWP gas injected into the system and can be implemented on the existing infrastructure~\cite{bib:Pinheiro2024_CO2Mixture_NIMA,bib:ATLAS_RPC_CO2mixture_NIMA2025}.

Studies of HFO/CO$_2$-based mixtures using iRPC-type double-gap prototypes at the GIF++ facility have shown that ecological candidates can reach plateau efficiencies above 95\% with cluster sizes comparable to the standard mixture~\cite{bib:CMS_HFOmixtures_NIMA2026}. In these tests the most promising mixture, containing about 35\% HFO-1234ze and 60\% CO$_2$, yields an effective mixture GWP of roughly 480, about a factor three lower than the present CMS mixture. The working point is higher than for the standard gas and the average gamma cluster charge is about 40\% larger, but the time resolution remains within the requirements for the CMS muon trigger. At the same time, HFO-based mixtures still require further aging studies, because their operation can lead to a higher production of HF$^-$ ions that may speed up detector degradation. In addition, many HFOs are part of the wider class of per- and polyfluoroalkyl substances (PFAS), very persistent compounds that are subject to increasing regulatory restrictions, so a full and permanent replacement of C$_2$H$_2$F$_4$ by HFOs remains an open question~\cite{bib:PFAS_Regulation_BlueFrog}.

A complementary line of work focuses on CO$_2$-based mixtures in which C$_2$H$_2$F$_4$ is only partially replaced. For the CMS iRPC system several such mixtures have been tested at GIF++, with CO$_2$ fractions between 30\% and 40\% and SF$_6$ concentrations up to 1\%~\cite{bib:Pinheiro2024_CO2Mixture_NIMA}. The selected candidates reduce the equivalent CO$_2$ emissions by about 15--26\% compared to the standard mixture while keeping the plateau efficiency close to 98\% in background-free conditions and above 90\% at gamma rates of order 1~kHz/cm$^2$. The muon cluster size stays around 1.7 strips and the working points are similar or even lower than for the standard gas. A first longevity campaign with an iRPC prototype operating at its working point in a CO$_2$-based mixture has accumulated about 40~mC/cm$^2$ of charge, corresponding to 4\% of the total charge expected by the end of CMS operation, with no significant change in resistivity, working point or efficiency~\cite{bib:Pinheiro2024_CO2Mixture_NIMA}. These results, obtained in an international effort with strong participation of the Brazilian RPC group, show that CO$_2$-based mixtures are a viable short- to mid-term solution for the CMS RPC upgrade programme, in case of problems with the recuperation system and/or when there is no access to leaky chambers.

\subsection {Brazilian person-power involved in the activities in the CMS-Muon-RPC project}
\label{subsec:cmspeople}

\begin{itemize}
\item[-] Alberto Santoro\footnote{Emeritus member}(UERJ)
\item[-] Dalmo Dalto\footnote{PhD student \label{refnotephd}}(UERJ)
\item[-] Dilson de Jesus Damião\footnote{Faculty\label{refnotefac}}(UERJ)
\item[-] Edmilson Manganote\footref{refnotefac}(CBPF/UNICAMP)
\item[-] Eduardo Alves Coelho\footnote{Researcher\label{refnoteres}}(CBPF)
\item[-] Eliza Melo Da Costa\footref{refnotefac}(UERJ)
\item[-] Fabio Marujo\footnote{Engineer}(CBPF)
\item[-] Felipe Silva\footref{refnotefac}(UERJ/UEA)
\item[-] Gilvan Augusto Alves \footnote{Permanent senior researcher}(CBPF)
\item[-] Helio Nogima\footref{refnotefac}(UERJ)
\item[-] Luiz Martins Mundim Filho\footref{refnotefac}(UERJ)
\item[-] Mapse Barroso Filho\footref{refnoteres}(CBPF)
\item[-] Maurício Thiel\footref{refnotefac}(UERJ)
\item[-] João Pedro Gomes Pinheiro\footref{refnotephd}(UERJ)
\item[-] Katherine Maslova Defante\footref{refnotephd}(UERJ)
\item[-] Kevin Mota Amarilo\footref{refnotefac}(UERJ)
\item[-] Raphael Gomes\footref{refnotephd}(UERJ)
\item[-] Sandro Fonseca de Souza\footref{refnotefac}(UERJ)
\item[-] Silas Santos De Jesus\footref{refnotephd}(UERJ)
\item[-] Wagner de Paula Carvalho\footref{refnotefac}(UERJ)
\item[-] Walter Luiz Aldá Junior\footref{refnotefac}(UERJ)
\end{itemize}


\subsection{Developments and possibilities of local activities among Brazilian institutions}

This proposal aims to equip the Nuclear and High Energy Physics Laboratories (LFNP) at UERJ and the COHEP at CBPF for experimental activities with gas detectors, particularly RPCs. For this reason, the main objective of this proposal is to enable the realization of CMS-RPC activities in the LFNP and CBPF, expanding the potential for instrumentation work. This will also allow studies of this and other gas detection technologies for future applications.

The development of the project is divided into two main stages. The first will include the installation of the gas-distribution and monitoring system necessary for the operation of gaseous detectors at the LFNP. This system comprises the infrastructure required to receive pre-mixed gas compositions, together with the controllers and gas distributors responsible for regulating pressure, monitoring consumption, validating gas tightness and ensuring stable delivery of the mixtures supplied to the RPC chambers. Six base gases will be employed in the studies, namely Ar, $CO_2$, $C_2H_2F_4$, iC$_{4}$H$_{10}$, SF$_6$ and HFO (HFO1234ze and/or HFO1233zd), although their individual high-purity cylinders and associated handling systems will be located at the CBPF, where the mixtures used in the LFNP studies will be prepared. This infrastructure will enable systematic investigations of RPC performance under different operational conditions, including current stability, avalanche formation, detector efficiency and the behaviour of the front-end electronics.

In parallel, the CBPF is establishing a dedicated Gas Mixing and Regeneration Laboratory within the INCT–CERN Brasil programme, which will serve as the national facility responsible for preparing all gas mixtures required for RPC studies. The six high-purity base gases foreseen in the project will be stored, handled and combined at CBPF using an automated high-precision mixing system equipped with mass-flow controllers, mass-flow meters and a diffusion chamber designed to guarantee homogeneity and stability of the final compositions. The mixtures prepared at CBPF will subsequently be supplied in cylinders to the LFNP and other Brazilian laboratories, where they will be used in RPC characterisation. In addition, CBPF, together with the Brazilian industry Recigases~\cite{bib:Recigases}, is developing a closed-loop gas-regeneration system capable of recovering, purifying and requalifying fluorinated mixtures from MARTA-type and CMS-RPC detectors. This combined infrastructure will enable comparative studies between fresh and regenerated mixtures, support investigations of eco-friendly gas alternatives, and strengthen the national R$\&$D programme on gaseous detectors in alignment with the objectives of the DRD1 Collaboration and the INCT–CERN Brasil initiative.

The first tests of this system will be carried out using a standard CMS-RPC chamber. By inserting known gas mixtures the detector response can be studied. This ensures good control of the entire system, from the mixture of gases, power supplies and the new digitizers. A high voltage source produces the electric field needed to produce signals in the RPC. Auxiliary detectors form a particle telescope that identify the passage of the particle, generating trigger for data acquisition.

In the second step, a prototype of the CMS iRPC chamber can be brought to the LFNP for characterization. Using other detectors for the trigger of cosmic muons, performance studies will be carried out on aspects such as (i) electric current as a function of high voltage, including intrinsic and avalanche components; (ii) avalanche scattering and its effects on spatial resolution and cross-talk; (iii) recovery time and dead time as a function of the applied high voltage; and (iv) time resolution.

At first, this work can be done with modular electronics, part of which is being acquired in this project. With preamplifiers connected to the pickup electrodes (strips) the reading will be done with the digitizers. Later, when the new Front End electronics of the CMS-RPC are available, the possibility of testing them in the LFNP and CBPF will be verified, together with the iRPC prototype. In this way, it will be possible to carry out comparative studies of signal production between the standard RPC and the iRPC of the CMS.

This same experimental arrangement can be used for alternative gases studies. This work should be done in conjunction with the CERN group, which has a facility for detector irradiation, GIF++. This feature makes it possible to study the generation and accumulation of impurities, produced in the electrical discharge process, which are converted into corrosive and harmful substances to the detector material. Mixtures that produce high concentrations of these substances are therefore undesirable and should be avoided. The existence of the testing infrastructure at the LFNP and CBPF  will allow collaborative work with the CERN group and the other institutions involved, dividing the performance study of different mixtures to each laboratory, in addition to allowing the verification of the redundancy of the results.


 %

\section{RPC Applications in Marta Experiment}

\subsection{Introduction}
MARTA (Muon Array with RPCs for Tagging Air showers) is a hybrid detector concept that combines the information from RPCs with data from another detector able to perform a calorimetric measurement of air showers \cite{MARTA_abreu2018}.
Was developed as an upgrade proposal for the Pierre Auger Observatory and which aims to improve the process of measuring the muonic component of  extensive air showers (EAS), which are created through the interaction of high-energy cosmic rays with the top of the Earth's atmosphere. At MARTA, the Resistive Plate Chambers (RPC) are placed below the Water-Cherenkov Detectors (WCD).

WCDs measure a combination of the muonic component and the electromagnetic signals from the air showers. As the electromagnetic signals are attenuated due to the water present in the WCD, the RPCs only perform the measurement of the muonic component, and through the calorimetric measurements allow, with a systematic control of uncertainties, a better understanding of the air showers, more specifically of the high-energy hadronic interaction properties. RPCs are widely used in the measurement of charged particles. If a particle crosses the detector when a high voltage is applied to the RPC, it will interact with the uniform electric field created in the gas gaps, generating an avalanche, which will induce a signal in the readout plane \cite{MARTA_luz2020}.

\begin{figure}[ht]
    \begin{center}
    \includegraphics[width=0.8\textwidth,keepaspectratio]{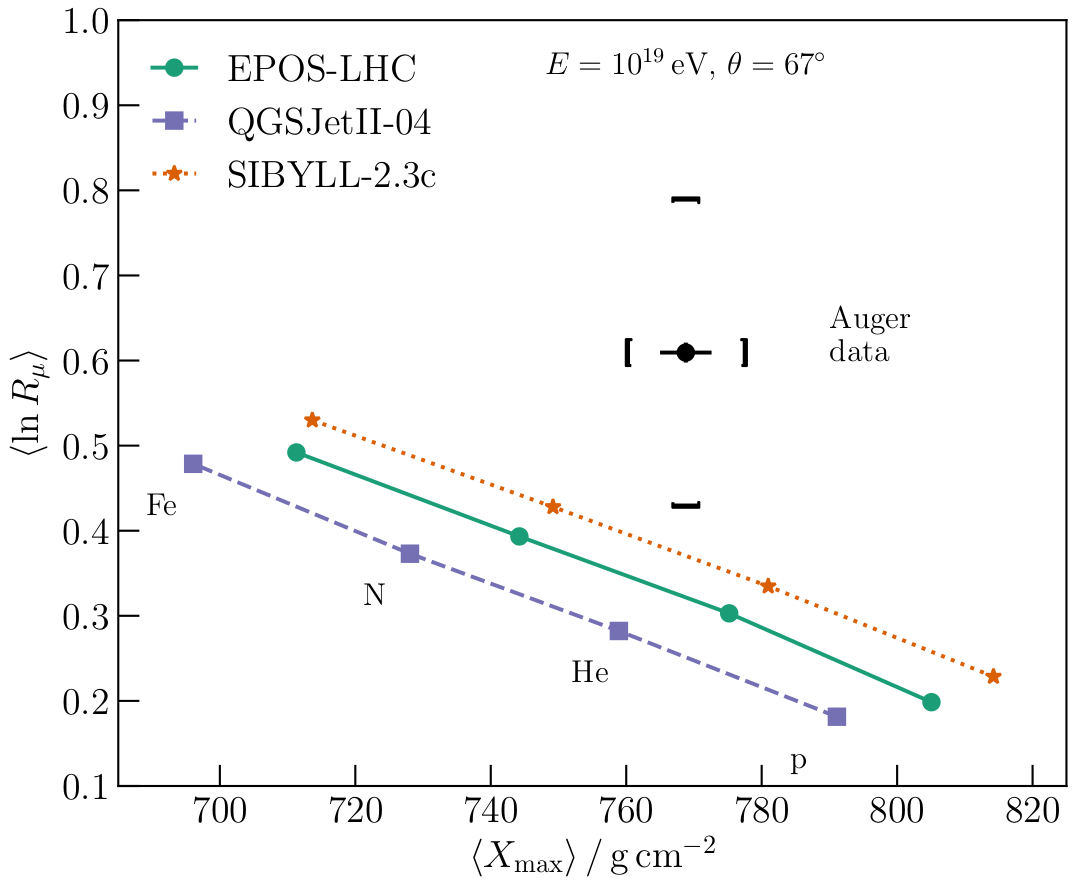}
    \end{center}
    \caption{The Pierre Auger Observatory results for the average logarithmic muon content $\langle{\ln{R_{\mu}}}\rangle$  as a fuction of the average maximum shower depth $\langle{X_{max}}\rangle$ at \;10$^{19}$ eV. The lines denote the distinct models predictions for different primary compositions. Taken from \cite{MARTA_ext_Aab_2021}.}
    \label{Marta_num_muons}
\end{figure}

The study of air showers allows us to study hadronic interaction at energy levels above those produced in particle colliders, and muons provide excellent links to hadronic interactions in EAS, as they are created in the decay of showers hadrons. Through recent simulations \cite{MARTA_ext_cazon2019}
it was found that the average muons distribution depends on the sum of the hadronic interactions and its shape mostly depends on what happens in the first iteration of the shower \cite{MARTA_tese_RJluz}.
However, there is a discrepancy between the simulations and the results obtained by the WCDs of the Pierre Auger Observatory, thus indicating the need to use specific detectors, in our case the MARTA project, to analyze the muon content. The (Figure \ref{Marta_num_muons}) shows this discrepancy. In the next section we will cover the structure and functioning of the RPCs modules, the conditions of use in outdoor environments and its ability to withstand adverse situations with a low cost of maintenance, gas and energy.

\subsection{The MARTA detector module}
The RPCs are used in a large number of experiments, but generally in indoor environments. However, as presented in \cite{MARTA_Lopes_2019},
there is a lot of evidence that they can be used in the most extreme environments with little maintenance and low energy and gas costs, making it more advantageous than other common detectors.
Therefore, MARTA RPCs are designed to operate in outdoor environments, with low power consumption, low gas flow and avalanche mode operation,  which in addition to providing them with better temporal resolution, is a less aggressive mode of operation for the detector, thus ensuring a long life to the detector.

\begin{figure}[ht]
    \begin{center}
    \includegraphics[width=0.8\textwidth,keepaspectratio]{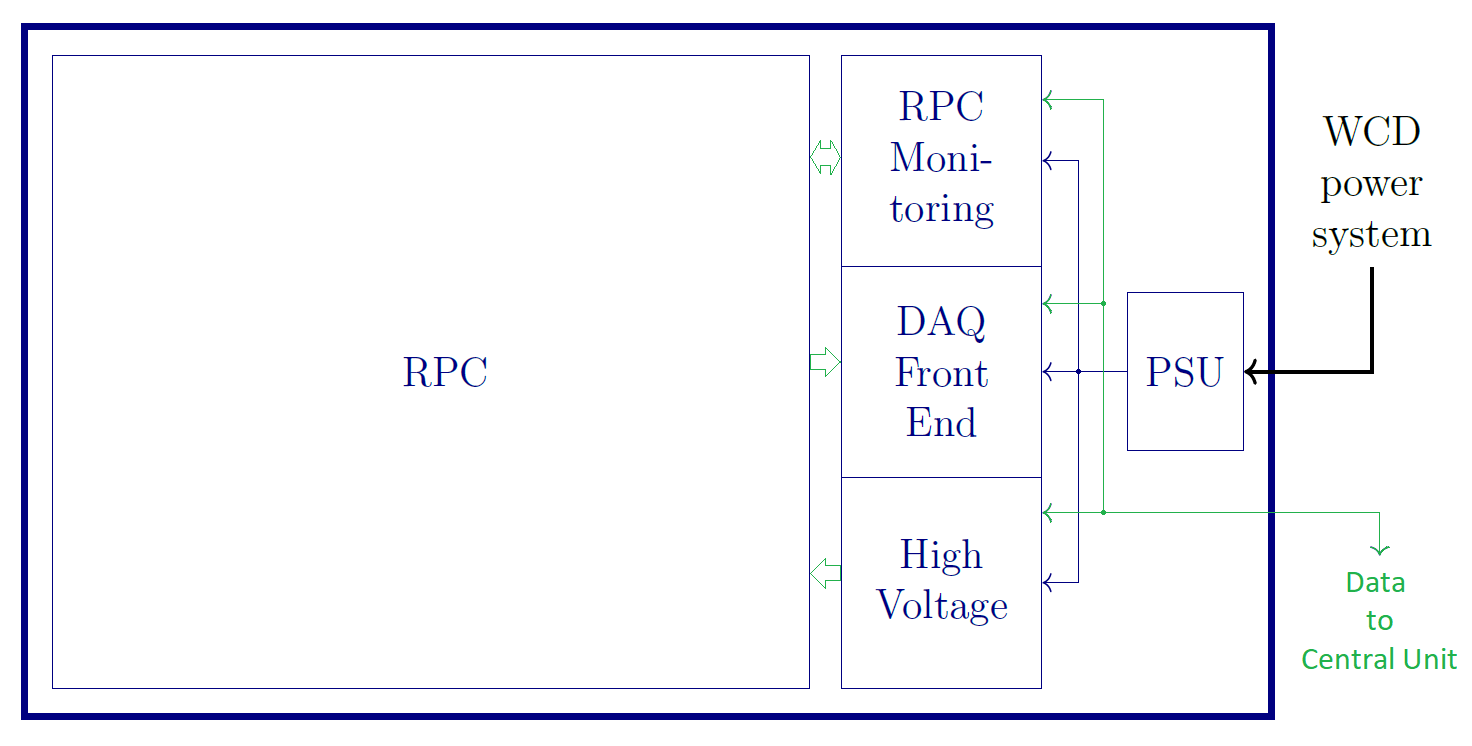}
    \end{center}
    \caption{RPC MARTA detector module .}
    \label{Marta_module_all}
\end{figure}

The RPC MARTA detector module (Figure \ref{Marta_module_all}) is mounted inside a hermetically sealed aluminum box, to thereby guarantee a second gaseous volume on the outside of the RPC acrylic gaseous volume. In order to thus lower possible contamination of the internal gas of the RPC, due to the fact that the acrylic is not completely sealed. A second aluminum area is added, where the data acquisition system (DAQ), the high voltage system and monitoring of HV, gas flow, atmospheric pressure, humidity and temperature at various points in the RPC are housed.

The detector proposal is composed of a module, with an RPC, which has a detection area of 1200x1500mm2 and with the following structure \cite{MARTA_abreu2018}:

\begin{itemize}
\item[-] A total of three resistive plates made of soda-lime glass, each 2 mm thick, mounted on top of each other;
\item[-] The resistive plates are separated by 1 mm gaps for the gas, making it a double gap chamber, filled with R-134a;
\item[-] The detector is glued to an acrylic box of 3 mm thickness;
\item[-] The readout plane is external and segmented in 8 × 8 pick-up electrodes (pads), each with dimensions 14 × 18 cm 2 and separated by a 1 cm guard ring;
\item[-] Coaxial cables transmit the signal induced in each pad to the DAQ.
\end{itemize}

\begin{figure}[ht]
    \begin{center}
    \includegraphics[width=0.8\textwidth,keepaspectratio]{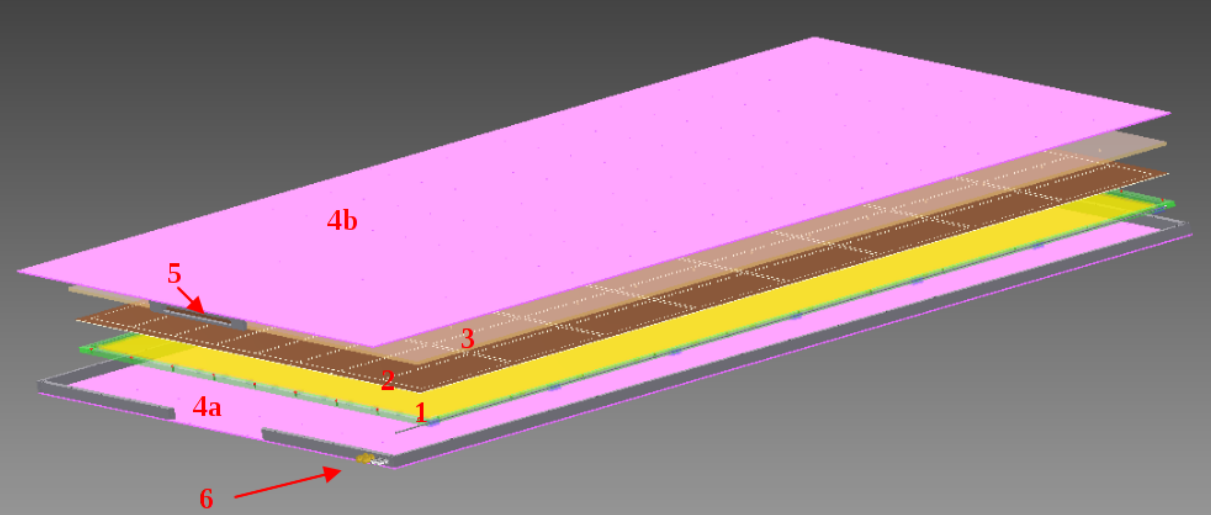}
    \end{center}
    \caption{Schematic drawing of the detector module: (1)120 × 150 cm$^{2}$ gas volume; (2)Readout pad plane 8 × 8 with 18 × 14 cm$^{2}$ area pads; (3)I$^{2}$C sensors layer of hydrophobic compressible foam; (4a)Aluminium case base; (4b)Aluminium case cover; (5)Signal feedthrough; (6)Gas and HV connectors.}
    \label{MARTA_schematic}
\end{figure}

In this detector (Figure \ref{MARTA_schematic}), the high resistivities of the plates prevent electrical discharges. The high-voltage electrodes are applied to the resistive plates, forming an intense uniform electric field, allowing the creation of avalanches through the passage of ionizing particles that come from the air showers. The readout of the signals is made by 8 × 8 pick-up electrodes pads and the signal is obtained from the center of the pad. Due to the physical separation between the readout electrodes and the gas volume, more complex readout patterns can be obtained than those collected by other acquisition systems. Small gas gaps provide fast detector response, maintaining good timescale resolutions and increasing detection efficiency.

In the next section, the implementation of the RPCs modules in the WCD will be discussed, it will also be explained about the attenuation that occurs in the electromagnetic signals that pass through the WCD tank and that later arrive at the RPCs.

\subsection{MARTA station}
\begin{figure}[ht]
    \begin{center}
    \includegraphics[width=0.4\textwidth,keepaspectratio]{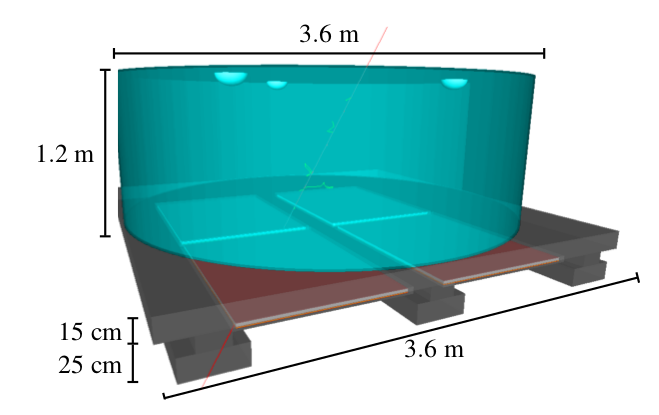}
    \end{center}
    \caption{MARTA station.}
    \label{MARTA_station}
\end{figure}

The proposal consists of four modules of RPCs that are designed to be placed below the water-Cherenkov detectors, in their concrete structure, therefore the dimensions of the RPCs are defined by the diameter of the detector tank. An example with all relevant dimensions are shown in (Figure~\ref{MARTA_station}).

In order for the cosmic rays to reach the RPCs, they need to pass through a 1.2m column of water, in addition to the 15cm of concrete that support the WCD, this process attenuates the electromagnetic component of the EAS. Therefore, the RPCs directly detect the muonic component of the air showers, which for the most part was also measured by the WCD. From this configuration, it is possible, by subtracting the muons measured by the RPC from the total signal obtained by the WCD, to obtain the electromagnetic component of the EAS \cite{MARTA_tese_RJluz}.

\begin{figure}[ht]
    \begin{center}
    \includegraphics[width=0.5\textwidth,keepaspectratio]{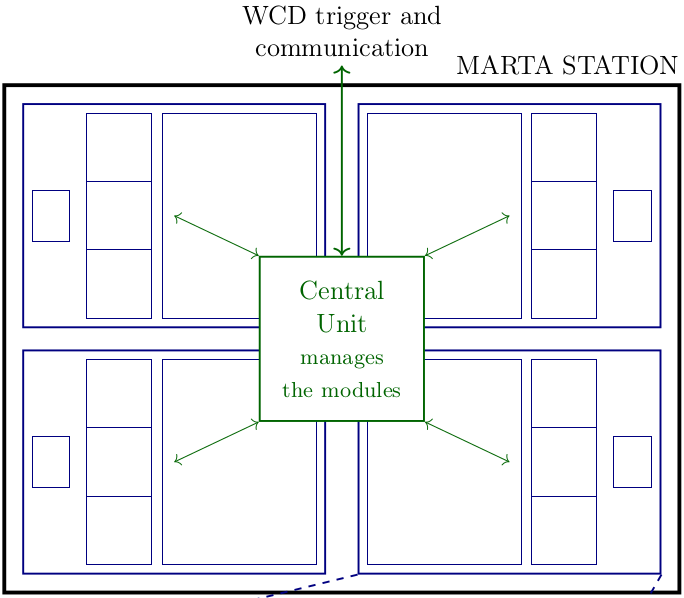}
    \end{center}
    \caption{Schematic of MARTA station.}
    \label{MARTA_schematic_station}
\end{figure}

The Figure \ref{MARTA_schematic_station} shows a schematic view of the MARTA station with the four RPCs modules. The MARTA station has a Central Unit responsible for the management of the modules and the interface with the WCD.

As there are no economically viable options, the HV board was made exclusively for the MARTA project. Its function is to control the applied HV dynamically. 
Consequently, the electric field between the gas gaps can be altered. As this quantity is sensitive to the temperature and pressure of the environment, the control of the electric field becomes fundamental, as the MARTA stations will be subject to the most adverse weather conditions at the Pierre Auger Observatory.

The monitoring system consists of a network of sensors inside the aluminum box. It's split between climate monitoring, gas monitoring and HV monitoring. Temperature, humidity and pressure are measured inside the RPC, which contains the sensitive volume. The power supply is given by the PSU. In the WCD the voltage used is 24 V, so the PSU is able to convert to the voltage required for each component of the module. The DAQ is a hybrid system capable of counting active RPC pads and measuring the charge induced in the detector.

The Central Unit has access to the data provided by all sensors and detectors in the MARTA station, acting as a data center. Also, it can control each of the modules independently.
In the next section we will discuss the design and operation of the MARTA engineering array, analyzing the location where it will be placed.

\subsection{MARTA engineering array Design}
\begin{figure}[ht]
    \begin{center}
    \includegraphics[width=0.4\textwidth,keepaspectratio]{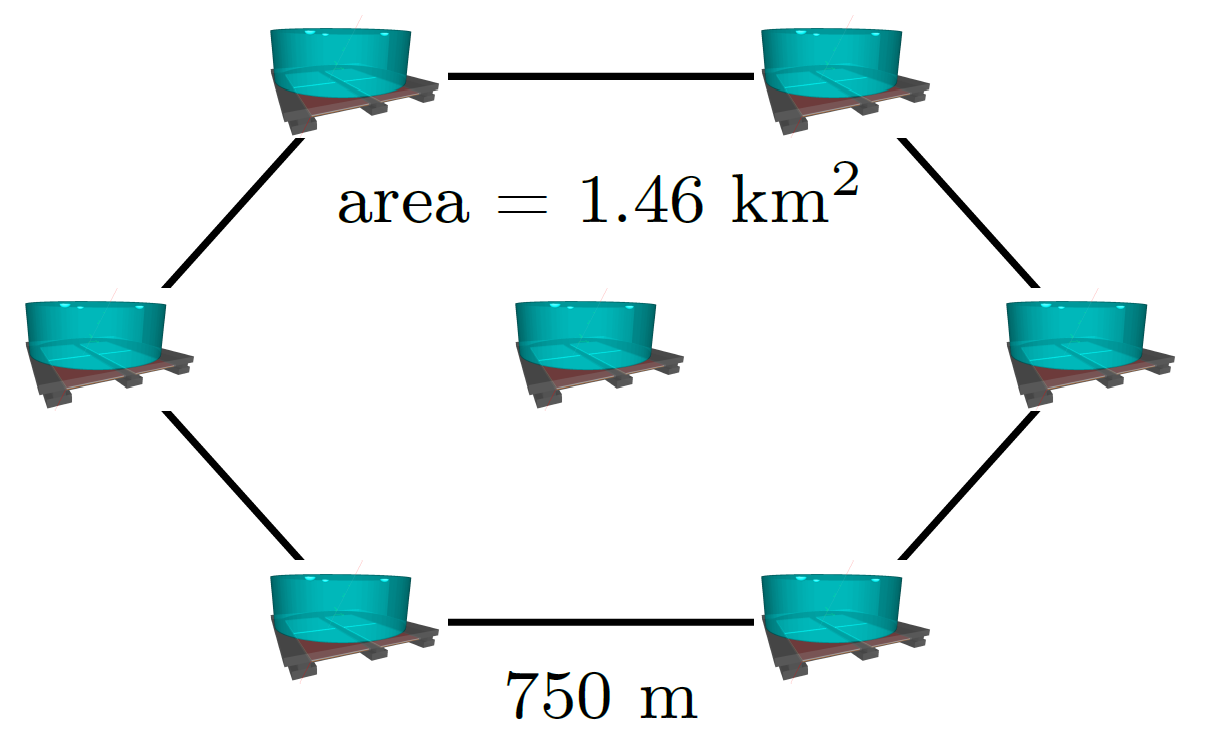}
    \end{center}
    \caption{MARTA engineering array.}
    \label{MARTA_engineering_array}
\end{figure}

\begin{figure}[ht]
    \begin{center}
    \includegraphics[width=0.4\textwidth,keepaspectratio]{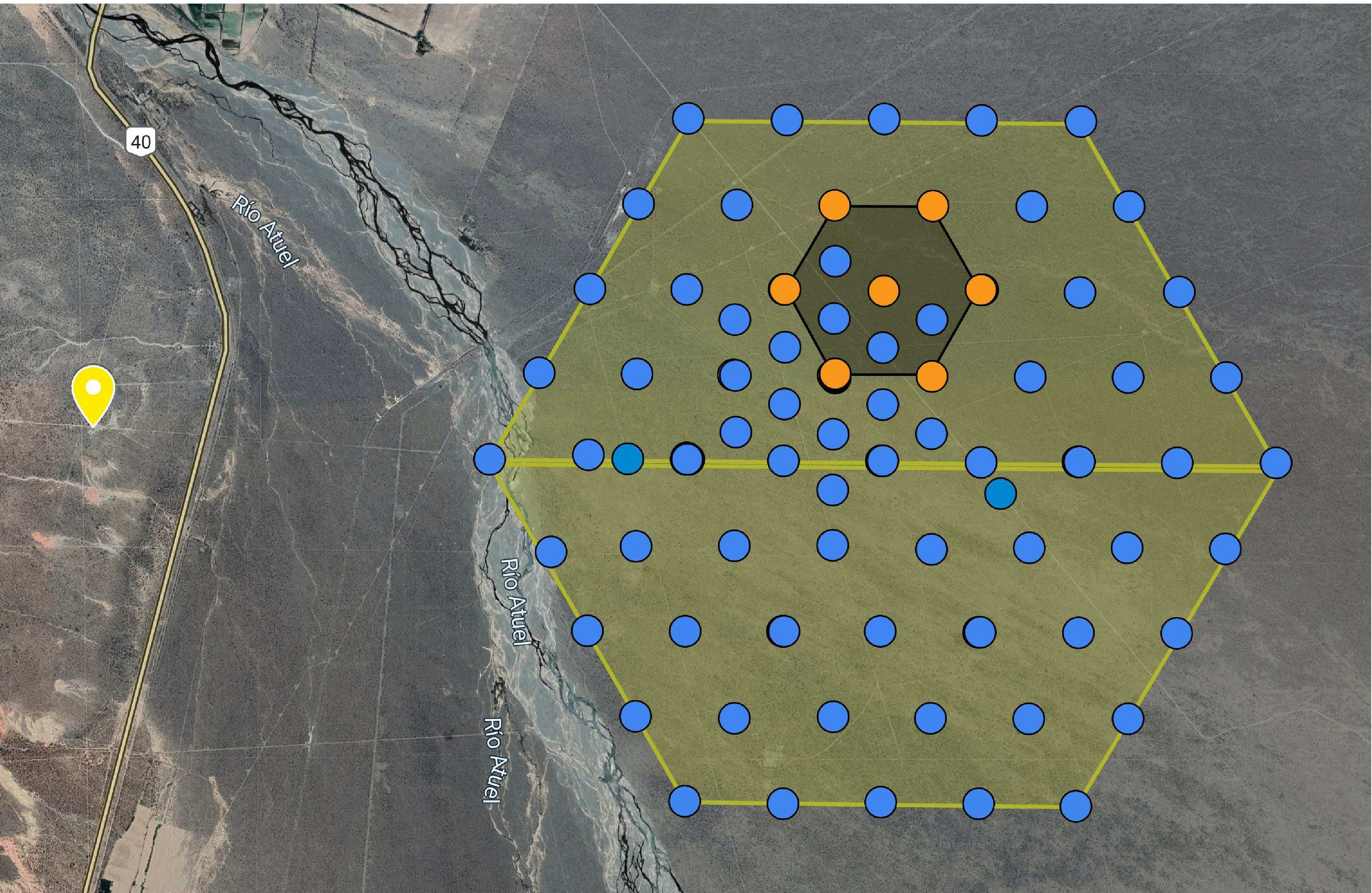}
    \end{center}
    \caption{MARTA engineering array in Auger Infill.}
    \label{MARTA_auger_infill}
\end{figure}

The proposed design for MARTA engineering array (Figure~\ref{MARTA_engineering_array}) is composed of six MARTA stations in a hexagon shape and with a MARTA station at its center. It will be installed on unitary cell in the observatory's Infill array (Figure~\ref{MARTA_auger_infill}), which has a higher density in terms of number of detectors, where the average distance between detectors is 0,75km, instead of the standard 1,5km. An array of this size will not be able to measure high energy events, but will serve as a basis for determining the calibration point for the muonic component of the EAS, thus allowing the calibration of the next detectors.

\newpage
\section{Possible applications of technology and involvement of companies and institutions}
\label{section_spinoffs}

\subsection{Applications of RPC detectors in imaging and time-of-flight systems}

As in many other cases of applied science, high-energy physics technologies can be adapted to address medical and industrial needs. Resistive Plate Chambers (RPCs), originally developed as tracking and timing detectors for particle physics, are a good example. Their simple and economical construction, reliable operation and very good time and position resolutions make them attractive for radiological imaging and nuclear medicine. The application of RPC technology~\cite{FONTE2000201} to positron emission tomography, known as RPC-PET~\cite{BLANCO200388}, is one successful example of this transfer of knowledge. Small-animal RPC-PET prototypes have demonstrated excellent time-of-flight performance and sub-millimetre intrinsic spatial resolution, while keeping the detector design relatively simple and cost effective~\cite{Razaghi_2019}.

To further improve the timing performance of RPCs, multigap configurations have been developed. In a multigap RPC, several thin gas gaps are placed in series between resistive plates. This increases the total gap thickness and the number of primary electrons produced by a traversing particle. The concept was introduced by M.~C.~S.~Williams and collaborators~\cite{CERRONZEBALLOS1996132}. Multigap Resistive Plate Chambers (MRPCs) can reach detection efficiencies above 95\,\% and time resolutions better than 100~ps, with moderate cost and scalable designs. Because of these features, MRPCs have become a standard technology for time-of-flight (ToF) systems in several particle physics experiments at the LHC and are being considered for future facilities such as the FCC~\cite{FCC:2018vvp}.

In Brazil, the expertise built in the CMS RPC collaboration has motivated national projects that explore RPC-based detectors in applied contexts. An example is the development of RPC-PET systems in collaboration with the Gaugit company~\cite{Gaugit}, which aims at PET scanners using locally developed technology and electronics. Other applications include RPC-based muon tomography systems for the non-destructive inspection of cargo containers~\cite{Baesso_2014}. In such systems, muons from cosmic rays are tracked before and after crossing the object under study, and the observed deflections are used to reconstruct its internal structure. These activities illustrate how the know-how acquired in large HEP experiments can generate spin-offs in medical imaging, security and industrial inspection.

\subsection{Spin-off project on RPC instrumentation and eco-gas mixtures at GIF++}

Building on the long-standing involvement of Brazilian groups in the CMS RPC collaboration, a dedicated spin-off project has been set up to focus on instrumentation with RPC-type gaseous detectors and on the study of alternative gas mixtures. The project is organised as a joint effort between UERJ and CERN and takes the HL-LHC operation of the CMS muon system as its main reference. Its goal is to deepen the understanding of RPC behaviour under high rate conditions, to validate eco-gas mixtures in realistic environments, and to consolidate Brazilian expertise in the construction, operation and monitoring of advanced RPC systems.

On the detector side, the project includes the assembly, operation and detailed characterisation of RPC and iRPC chambers equipped with modern front-end electronics. Chambers are studied under different gas mixtures, including standard C$_2$H$_2$F$_4$-based compositions and lower-GWP alternatives. Performance measurements cover efficiency, cluster size, timing and induced charge, together with the analysis of signal waveforms to understand how the gas composition and operating conditions affect the avalanche development. These studies are carried out both in surface laboratories and in the CMS environment, in close connection with the operation of the installed RPC and iRPC systems.

A central element of the project is the use of the Gamma Irradiation Facility (GIF++) at CERN~\cite{gifpp} to test RPC chambers under intense gamma backgrounds representative of HL-LHC conditions. A dedicated cosmic-ray trigger system is being designed, installed and commissioned at GIF++ so that RPCs can be operated at a fixed angle to cosmic muons during LS3, when no accelerator beams are available. This setup will allow long-term performance and aging studies with eco-gas mixtures, including continuous monitoring of detector currents, efficiencies and working points. In parallel, a laboratory at UERJ is being prepared for RPC assembly, gas studies and signal analysis, ensuring that the methods and tools developed at CERN are transferred to Brazil and remain available for future R\&D and applications beyond the CMS experiment.

\subsection{Studies of signal production in RPC and the relationship between waveform and avalanche process development}

This research project has as its main objective to develop advanced methods for the analysis of signals from particle detectors, aiming both at the quantification of quantities of interest and their use in probes for internal physical processes. Signals from RPC will be investigated in the context of the search for alternative gas mixtures that have a lower environmental impact than those currently in use. This topic is highly relevant to HEP experiments, which employ such detectors on a large scale. Measurements of parameters such as efficiency and temporal and spatial resolution are generally compared with those of currently used mixtures. However, some effects due to molecular composition may be reflected in a way that is not perceptible by analyzing these measurements. The study of the signal waveform can be an important ally in understanding important effects that different mixtures may present. In this work, we focus on two measurement fronts: 
\begin{itemize}
    \item[-] Measurements of parameters related to the signal size, such as pulse height and electric charge, in addition to temporal characteristics, such as rise time and pulse width;
    \item[-] Analysis of the relationship between the signal shape and the underlying physical processes, using machine learning methods.
\end{itemize}

The study will include experimental measurements carried out both in the LFNP/UERJ and at CERN, also including the development of machine learning algorithms and computational simulations. Although the focus is on the analysis of RPC signals, the methods developed have potential for application in other technologies, given the rapid advances in the area of signal processing driven by artificial intelligence.


\bibliographystyle{unsrtnat}

\bibliography{sample}


\end{document}